\def\thetac{\theta_C}

\def\dsone{|\Delta S|=1}

\def\smalllinestretch{\renewcommand{\baselinestretch}{0.7}}
\def\largelinestretch{\renewcommand{\baselinestretch}{1.0}}
 \def\tr {\, \mbox{tr} \,}

 \def\im {\, \mbox{Im} \,}
 \def\re {\, \mbox{Re} \,}
 
 \def\A{{\cal A}}
 \def\B{{\cal B}}
 \def\L{{\cal L}}

 \def\O{{\cal O}}

 \def\delg{\Delta g(K^\pm\to 3\pi)}
 \def\gt{\widetilde{G}}

\newcommand{\su}[1]{\mbox{\small\it #1}}
\documentstyle{article}

\def\largelinestretch{\renewcommand{\baselinestretch}{1.2}}
\largelinestretch\small\normalsize
\title{
        On the $p^4$--corrections to $K \to 3\pi$ decay amplitudes\\
        in nonlinear and linear chiral models
      }
 \author{
A.A.Bel'kov${}^1$,
G.Bohm${}^2$,
A.V.Lanyov${}^1$,
A.Schaale${}^3$
\\
\\
\small
${}^1$
        Particle Physics Laboratory, Joint Institute for Nuclear Research,
\hfill\\
\small
        Head Post Office, P.O. BOX 79, 101000 Moscow, Russia
\hfill\\
\small
${}^2$
        DESY--IfH Zeuthen, Postfach 15735 Zeuthen, Germany
\hfill\\
\small
${}^3$  TRIUMF, 4004 Wesbrook Mall, Vancouver, B.C., Canada V6T 2A3
\hfill\\
\small (supported by Deutscher Akademischer Austauschdienst, DAAD)
}
\begin{document}
\largelinestretch\normalsize
   \thispagestyle{empty}
   \begin{titlepage}
   \thispagestyle{empty}
   \maketitle
   \begin{abstract}

    The calculations of isotopic amplitudes and their results for the
direct $CP$--violating charge asymmetry in $K^\pm \to 3\pi$ decays within
the nonlinear and linear ($\sigma$--model) chiral Lagrangian approach
are compared with each other.
    It is shown, that the latter, taking into account intermediate scalar
resonances, does not reproduce the $p^4$--corrections of the nonlinear
approach introduced by Gasser and Leutwyler, being saturated mainly by
vector resonance exchange.
    The resulting differences concerning the $CP$ violation effect are
traced in some detail.

   \end{abstract}
   \end{titlepage}

   The purpose of this short note is to clarify further the
model--dependence of various predictions concerning the manifestation
of direct $CP$ violation in the charge asymmetry of $K^\pm \to 3\pi$ decays.
   Estimates for this charge asymmetry have been given in the soft
pion limit \cite{CPsoft-pion-old}, resulting in rather small effects.
   Only after taking into account higher orders of chiral perturbation
theory, a large value for $|\Delta g|$ in relation to
$\re ({\varepsilon}'/\varepsilon)$ has been derived in \cite{CPour-prev},
which has been met with some criticism
[3--6]
(see also the discussion on the Joint Lepton--Photon and Europhysics
Conference, Geneva 1991 \cite{geneva}).
   After a reformulation of the bosonization prescription
\cite{bosoniz-our}, and a more detailed investigation of the origin
for the enhancement, a (slightly corrected, see below) new result for
$|\Delta g|/\re ({\varepsilon}'/\varepsilon)$ has been given in
\cite{CPour-last}.

   The effects of $CP$-violation to be observed appear from interfering
amplitudes with different quantum numbers.
     In $K^{\pm} \rightarrow 3\pi$ decays (as opposed to
$K^{\pm} \rightarrow 2\pi$, where only two amplitudes with isospin changes
$|\Delta I| = 1/2$ and $|\Delta I| = 3/2$ interfere) there is possible an
additional contribution from the interference of two different amplitudes
both with $|\Delta I| = 1/2$.
     In the soft--pion limit this additional contribution becomes zero,
and only interferences of amplitudes with $|\Delta I|=1/2$ and $3/2$ can
contribute to the  charge asymmetry $\Delta g$ in this limit.
     However taking into account $p^4$-corrections and rescattering of mesons
strongly modifies the soft-pion amplitudes and leads to a large value for this
contribution, increasing the charge $CP$-asymmetry in
$K^{\pm} \rightarrow 3\pi$ compared to old estimates in the soft-pion
approximation \cite{CPsoft-pion-old}.

   In view of the great importance of possible direct $CP$--violation
effects other than those intensively investigated in $K^0_{L,S}$
decays, it is certainly worthwhile to compare the prediction with
those found in other models and to trace possible differences.
   In recent papers \cite{shabalin2}, Shabalin investigated the
charge asymmetry in $K^{\pm} \rightarrow 3\pi$ decays in the framework
of a linear $\sigma$--model, using the same ansatz for the weak
interaction Lagrangian on quark level \cite{vzsh} as used in our
papers \cite{CPour-prev,CPour-last}.
   The result of \cite{shabalin2} differs from ours \cite{CPour-last}
by a factor $10 \div 20$.
   The origin of this discrepancy can be traced, by a straight forward
comparison of both calculations, to the different treatment of the
higher order corrections in both models, despite their practical
equivalence with respect to the description of other data on
$K$--decays, and rough numerical agreement in many intermediate
parameters.
   We shall not enter a discussion of the absolute size of
$CP$--violating effects, which have been investigated in detail by
\cite{CP-buras}, restricting ourselves to a consideration of the
relation between the charge asymmetry in $K^{\pm} \rightarrow 3\pi$ and
$\re ({\varepsilon}'/\varepsilon)$ as measured in
$K^0_{L,S} \rightarrow 2\pi$ experiments.

   The effective Lagrangian describing nonleptonic weak interactions
with strangeness change $\dsone{}$ is given on the quark level by
\cite{vzsh,gilman-wise}:
 \begin{equation}
 \L^{\su{nl}}_{\su{w}} = \gt \sum_{i=1}^6 c_i \, {\cal O}_i \; .
 \label{weak-lagr}
 \end{equation}
   Here $ \gt = \sqrt2 \, G_F \, \sin \thetac \,\cos \thetac$ is the
weak coupling constant; $c_i$ are Wilson coefficient functions;
$\O_i$ are the four-quark operators consisting of products of left-
and/or right-handed quark currents. In the present paper we will use the
operators $\O_i$ in the representation of Shifman--Vainshtein--Zakharov
\cite{vzsh}.
   The bosonized Lagrangian of nonleptonic four--quark weak interactions
(\ref{weak-lagr}) and the corresponding meson currents can be obtained
by the functional method using the generating functional for Green
functions of quark currents introduced in \cite{pich-rafael} and
\cite{bosoniz-our}.
    In such an approach the quark determinant, which leads to the effective
Lagrangian of meson strong interaction, generates also the meson
currents and scalar densities entering in the bosonized version of the
nonleptonic weak Lagrangian.

    The corresponding nonlinear effective meson strong Lagrangian, including
$p^2$-- and higher order derivatives terms, can be presented in the following
general form
\begin{eqnarray}
{\cal L}^{(nlin)}_{eff}&=&-\frac{F^2_0}{4} \tr \big( L_{\mu} L^{\mu} \big)
                          +\frac{F_0^2}{4} \tr \big[ M(U + U^+) \big]
\nonumber \\
&&+ \bigg( L_1-\frac{1}{2}L_2 \bigg)\,\big( \tr L_{\mu} L^{\mu}\big)^2
  + L_2 \tr \bigg(\frac{1}{2}[L_{\mu},L_{\nu}]^2+3(L_{\mu}L^{\mu})^2\bigg)
  + L_3 \tr \big( (L_{\mu} L^{\mu})^2 \big)
\nonumber \\
&&+ L_4 \tr \big( D_{\mu}U\,\overline{D}^{\mu}U^+ \big)\,
        \tr M\big( U + U^+ \big)
  + L_5 \tr D_{\mu}U\,\overline{D}^{\mu}U^+\big( MU + U^+M \big)
  + ...\,\,,
\label{LeffGL}
\end{eqnarray}
where the  dimensionless structure constants $L_i$ were introduced by Gasser
and Leutwyler in ref.\cite{Gasser-Leutwyler}.
Here we write down only the $p^2$-- and $p^4$--terms relevant for the
description of nonleptonic kaon decays restricting ourselves to terms
up to order $m_0$ in the quark mass.
    $F_0=89 MeV$ is the bare $\pi$ decay constant;
$L_{\mu}=D_{\mu}U\,U^+ $, where $U$ is a pseudoscalar meson matrix;
\begin{eqnarray}
D_{\mu}* = \partial_{\mu}* + (A^{(-)}_{\mu}* - *A^{(+)}_{\mu})\,,
\quad
\overline{D}_{\mu}* = \partial_{\mu}* + (A^{(+)}_{\mu}*-*A^{(-)}_{\mu})
\label{deriv}
\end{eqnarray}
are the covariant derivatives; $A^{(\pm)}_{\mu} = V_{\mu} \pm A_{\mu}$
with $V_{\mu}$, $A_{\mu}$ being the external vector and axial--vector fields.
    The meson mass matrix $M = diag(\chi^2_u,\chi^2_d,\chi^2_s)$ with the
parameters
$\chi^2_i = -2m^i_0\!\!<\!\!\overline{q}q\!\!>\!\!F^{-2}_0$,
where $<\overline{q} q >$ is the quark condensate, can be fixed by the
spectrum of pseudoscalar mesons.
   The coefficients $L_i$ are given by $L_1-\frac{1}{2}L_2 = L_4 = 0$ and
\begin{eqnarray}
  L_2 = \frac{N_c}{16 \pi^2}\frac{1}{12}\,,\quad
  L_3 = -\,\frac{N_c}{16 \pi^2}\frac{1}{6}\,,\quad
  L_5 =  \frac{N_c}{16 \pi^2}x(y-1)\,,
\label{Lcoeff}
\end{eqnarray}
where $y = 4\pi^2 F^2_0/(N_c \mu^2)$ and
$x = -\mu F_0^2/(2 <\! \! \overline{q} q \! \! >)$, $\mu = 380 MeV$ is
the averaged constituent quark mass.

    The pseudoscalar meson matrix $U$ arises in a nonlinear
parameterization of chiral symmetry from the following representation
of the combination of the external scalar (S) and pseudoscalar (P) fields:
\begin{eqnarray*}
               \Phi = S + iP = \Omega \,\Sigma \,\Omega \,.
\end{eqnarray*}
Here $\Sigma(x)$ is the matrix of scalar fields belonging to the
diagonal flavour group while the matrix $\Omega(x)$ represents the
pseudoscalar degrees of freedom $\varphi$ living in the coset flavour space
$U(3)_L \! \times \! U(3)_R/U_V(3)$, which can be parameterized by
the unitary matrix
\begin{eqnarray*}
  \Omega (x) = \exp \left(\frac{i}{\sqrt{2}F_0} \varphi (x) \right)\,,
\quad
  \varphi (x) = \varphi^a(x)\frac{\lambda^a}{2}\,,
\end{eqnarray*}
where $\lambda^a$ are the generators of the $SU(3)$ flavour group.
   Assuming approximate flavour symmetry of the condensate
($\Sigma \approx \mu \bf 1$) one obtaines $\Phi = \mu \Omega^2 = \mu U$
with $U=\Omega^2$.

    The bosonized $(V \mp A)$ meson currents, corresponding to the quark
currents $\overline{q}\frac{1 \mp \gamma_5}{2}
\gamma_{\mu}\frac{\lambda^a}{2}q$, can be obtained by varying the
quark determinant with redefined vector and axial--vector fields
$$
V_{\mu}  \to V_{\mu}-i(\eta_{L\mu}+\eta_{R\mu})\,,\quad
A_{\mu}  \to A_{\mu}+i(\eta_{L\mu}-\eta_{R\mu})\,,
$$
over the external sources
$\eta_{L,R\mu}= \eta_{L,R\mu}^a \frac{\lambda^a}{2}$ coupling to the
corresponding quark currents \cite{bosoniz-our} (the fields $V_{\mu}$,
$A_{\mu}$ enter in the covariant derivatives).
    In this way the effective Lagrangian (\ref{LeffGL}) generates the
bosonized $(V-A)$ meson current of the form
\begin{eqnarray}
J^{(nlin) a}_{L\mu}&=&i\frac{F^2_0}{4}\tr \Big(\lambda^a L_{\mu}\Big)
\nonumber \\
&& -i\tr \Big\{ \lambda^a \Big[
      2 L_2 L_{\nu}L_{\mu}L^{\nu}
   + (2 L_2 + L_3) \big\{ L_{\mu},L_{\nu}L^{\nu}\big\}
\nonumber \\
&& - \frac{1}{2} L_5 U^+\big\{(MU+U^+M),L_{\mu}\big\}U
         \Big] \Big\}\,.
\label{JeffGL}
\end{eqnarray}
    This current determines the meson matrix elements of
$|\Delta I| = 1/2$ ($\O_{1,2,3}$) and $|\Delta I| = 3/2$ ($\O_4$)
non--penguin four--quark operators consisting of products of left--handed
quark currents.
    The first term in (\ref{JeffGL}) is generated by the kinetic term
of the Lagrangian (\ref{LeffGL}) while all other terms originate from its
$p^4$--part.

    The $(S-P)$ meson current corresponding to the bosonized scalar
density and generated by the Lagrangian (\ref{LeffGL}) can be obtained
by variation of the quark determinant with redefined scalar and
pseudoscalar fields
 \begin{eqnarray}
S \to S - (\eta_L +\eta_R)\,,\quad P \to P - i(\eta_L - \eta_R) \,,
\label{SPredif}
\end{eqnarray}
where ${\eta}_{L,R} = \eta_{L,R}^a \frac{\lambda^a}{2}$ are the external
sources coupling to the the quark densities
$\overline{q} \frac{1 \mp \gamma_5}{2} \frac{\lambda^a}{2}q$.
     The corresponding $(S-P)$ meson current has the form
\begin{eqnarray}
J^{(nlin) a}_L &=& \frac{F^2_0}{8\mu}\tr \big(\lambda^a \partial^2U \big)
   + \frac{F^2_0}{4}\mu R \tr \big(\lambda^a U \big)
\nonumber \\
&& - \frac{1}{\mu} \tr \Big\{ \lambda^a \Big[
     L_2\partial_{\mu}\big( L_{\nu}L^{\mu}L^{\nu} \big)
   + \big( 2L_2+L_3 \big) \partial_{\mu}\big( L_{\nu}L^{\nu}L^{\mu}\big)
\nonumber \\
&& -\frac{1}{2} L_5 \Big( \partial_{\mu}\big( (MU+U^+M)L^{\mu}U \big)
                         + 2\mu^2 R L_{\mu}L^{\mu} \Big) \Big] \Big\}\,,
\label{JSPnred}
\end{eqnarray}
where $R=<\!\!\overline{q}q\!\!>\!\!/(\mu F^{2}_0)$.
    Here, the first and second terms are generated by the kinetic and mass
terms of the Lagrangian (\ref{LeffGL}), respectively, while all other terms
originate from its $p^4$--part.

     In papers \cite{shabalin1,shabalin2} the $K \to 3\pi$ decays amplitudes
were calculated within the linear $\sigma$--model with broken
$U(3)_L \! \times \! U(3)_R$ symmetry.
     The effective Lagrangian of meson strong interactions used in
ref.\cite{shabalin2} is of the form
\begin{eqnarray}
{\cal L}^{(lin)}_{eff}&=&
    \frac{1}{2}\tr \big({\partial}_{\mu} \Phi \, {\partial}^{\mu} \Phi^+ \big)
  + \frac{F_0}{2 \sqrt{2}} \tr \big[\big( \Phi + \Phi^+ \big)\widetilde{M}\big]
\nonumber \\
&&- c\tr \big( \Phi \, \Phi^+ - A^2\lambda^2_0 \big)^2
  - c\xi \Big( \tr \big( \Phi \, \Phi^+ - A^2\lambda^2_0 \big) \Big)^2 \,,
\label{Lefflin}
\end{eqnarray}
where $\Phi = \hat{\sigma} + i\hat{\pi}$; $\hat{\sigma}$ and
$\hat{\pi}$ are matrices of nonets of scalar and pseudoscalar mesons;
$\lambda_0 = \frac{1}{\sqrt{3}} \bf 1$; $\widetilde{M}$ is the mass
matrix, which gets in the approximation $m^0_u = m^0_d$ the form
$$
  \widetilde{M} = \frac{1}{\sqrt{3}}\big( 2m_K^2 + m^2_{\pi}\big) \lambda_0
                 -\frac{2}{\sqrt{3}}\big( m_K^2 - m^2_{\pi}\big) \lambda_8 \,.
$$
    The parameter $c$ can be expressed through the masses $m_{\pi}$ and $m_K$
and the $\pi , K \to \mu \nu$ decay constants $F_{\pi , K}$:
$$
   c = \frac{m_K^2 - m^2_{\pi}}{4F^2_\pi (F_K/F_\pi -1)(2F_K/F_\pi -1)}\,;
$$
the constant $A$ is connected to the quark condensate, and the value of the
parameter $\xi = -0.225$ is fixed from the $K_{e4}$ decay form factors.
    The $(V-A)$ and $(S-P)$ meson currents originating from kinetic
and mass terms respectively of the effective Lagrangian
(\ref{Lefflin}) and used in \cite{shabalin2} are
\begin{eqnarray}
J^{(lin) a}_{L\mu} = i\tr \big(\lambda^a \partial_{\mu} \Phi \, \Phi^+ \big)\,,
\label{JVAlin}
\end{eqnarray}
\begin{eqnarray}
J^{(lin) a}_L = \frac{\sqrt{2} F_0 m^2_{\pi}}{m_u + m_d}
                \tr \big(\lambda^a \Phi \big)\,.
\label{JSPlin}
\end{eqnarray}

    The penguin diagrams give a contribution to the effective weak
interaction proportional to the $|\Delta I| = 1/2$ operator
\footnote { The contribution of the operator $\O_6$ is small and is therefore
neglected .}
\begin{eqnarray*}
{\O}_5 = \bar{d}_L \gamma_\mu \lambda^a_c s_L
  \left( \sum_{q=u,d,s} \bar{q}_R \, \gamma^\mu \, \lambda^a_c \, q_R \right)
  \stackrel{Fierz}{\longrightarrow}
                           -4\sum_{q} \bar{d}_L\, q_R \cdot \bar{q}_R\, s_L .
\end{eqnarray*}
     We can find all the meson matrix elements of $\bar{q}_L\, q'_R$,
for example, using a modified version of the $QCD$ Lagrangian in which
the quarks are coupled to external $U(3)_L \times U(3)_R$ gauge fields, and
the quark mass term is replaced by
$$
   \sum_{q} m^0_q \bar{q}\,q \rightarrow
   \sum_{q,q'} \kappa_{qq'}(x) \bar{q}_L\, q'_R + h.c.\,,
$$
where $\kappa(x)$ is an arbitrary space-time dependent $3\times 3$ matrix of
external fields (see the detailed discussion in \cite{chivikula}).
     In this approach the quark mass $m^0_q$ is replaced by the scalar source
$\kappa_{qq'}(x)$ for the quark density.
     The meson matrix element of the operator $\O_5$ is found by replacing
the quark density $J_L^{(q)}=\bar{q}_L\, q'_R$ by the meson
scalar current $J_L^{(m)}=\frac{1}{4} <\bar{q}\,q> U_{q'q}^+$
generated from the chiral symmetry breaking part of the meson Lagrangian.
     Then
$$
      <{\cal O}_5>_{mes} = -\frac{1}{4} <\bar{q}\,q>^2
                            \sum_{q} U_{sq} U_{qd}^+ = 0
$$
because $UU^+ = 1$, therefore $(UU^+)_{sd} = 0$.
     This poses a problem for the naive penguin treatment in the chiral
Lagrangian language: on quark level in the simple vacuum incertion
approximation the meson matrix element $\O_5$ does not disappear.

     To solve this problem, in ref.\cite{chivikula} the new additional
symmetry breaking term $\sim tr \,m_0 D^2 U$ was added to the chiral symmetry
breaking part of the effective meson Lagrangian (\ref{LeffGL}).
     This new term leads  to nonzero meson matrix elements of the penguin
operator ${\cal O}_5$ due to the appearence of the additional contribution
to the scalar density $\sim tr \, {\partial}^2 U$.
     This additional contribution automatically arises in \cite{bosoniz-our}
from the kinetic term of the effective Lagrangian (\ref{LeffGL}) via the
replacement $\mu U^+ \rightarrow \mu U^+ - 2{\eta}_L (x)$, corresponding
to the redefinition of the scalar and pseudoscalar external fields
(\ref{SPredif}).
     The term $\sim tr \, {\partial}^2 U$ in the bosonized scalar density
was used in \cite{buras1,paschos} and in our calculations
\cite{CPour-prev,CPour-last}.
    Concerning the problem of the chiral bosonization of penquin
operators in the linear $\sigma$--model, one should pay attention to the
fact that in this case $(\Phi \Phi^+)_{sd} \neq 0$ and the $(S-P)$
current (\ref{JSPlin}), generated by the mass term of Lagrangian
(\ref{Lefflin}), already ensures the nonzero value of the meson matrix
element $\O_5$.
    Nevertheless, it is obvious that the contribution
$\sim tr \, {\partial}^2 \Phi$ to the $(S-P)$ current must arise also
in the same way from
the kinetic part of Lagrangian (\ref{Lefflin}) after redefinition of
$\hat{\sigma}$ and $\hat{\pi}$ fields.
    However, the corresponding contribution was not considered in
ref.\cite{shabalin2}.

    The $K^+ \to 3\pi$ decay amplitudes can be parametrized using isospin
relations as \cite{devlin}
\begin{eqnarray}
T_{K^+ \to \pi^+ \pi^+ \pi^-} &=& 2 \, (\A_{11}+\A_{13})
                                 -Y \, (\B_{11}+\B_{13}-\B_{23}) + O(Y^2),
\nonumber \\
T_{K^+ \to \pi^0 \pi^0 \pi^+} &=& (\A_{11}+\A_{13})
                                 +Y \, (\B_{11}+\B_{13}+\B_{23}) + O(Y^2),
\end{eqnarray}
   where $Y = (s_3-s_0)/m_\pi^2$ is the Dalitz variable and
$s_i={(k-p_i)}^2$, $s_0=m_K^2/3 + m_\pi^2$;
$k$, $p_i$ are four-momenta of the kaon and $i$th pion ($i=3$ belongs
to the odd pion).
   The Dalitz-plot distribution can be written as a power series
expansion of the amplitude squared, $|T|^2$, in terms of the
corresponding kinematical variables $Y$ and $X$
$$
|T|^2 \propto 1 + gY + ...
$$

   The isotopic amplitudes $\A_{IJ}$, $\B_{IJ}$ of $K\to3\pi$
decays have two indices: $I$, the isospin of the final state, and
$J$, the doubled value of isospin change between the initial and
final states.
   It is customary in anology to the $2\pi$-system to introduce strong
phase shifts $\alpha_1$, $\beta_1$ and $\beta_2$ corresponding to the
relevant isospin states $I=1_s$ (symmetric), $I=1_m$ (mixed symmetric),
$I=2$ by writing
\[ \A_{11} + \A_{13} = (a_{11} + a_{13}) \, e^{i\alpha_1}, \qquad
   \B_{11} + \B_{13} = (b_{11} + b_{13}) \, e^{i\beta_1}, \qquad
   \B_{23} = b_{23} \, e^{i\beta_2}.
\]
   We shall use this representation here only in order to display more
cleary the relationships between the main contributions to the direct
CP-violation effect and for the comparision with calculations in other
papers.
   Because the strong Hamiltonian is not necessarily diagonal with
respect to the $I=1_s, I=1_m$ isospin states and, if isospin breaking
is included, even $I=1$ and $I=2$ states get mixed, leading to the
necessity of introducing more phases, the exact
calculations of $\delg$ have to be done using the
complex quantities $\A_{IJ}$, $\B_{IJ}$ given below by (\ref{defab})
directly, without introducing the strong phases $\alpha_1$,
$\beta_{1,\, 2}$ explicitly.

   Let us next introduce the contributions of the four-quark
operators $\O_i$ to the isotopic amplitudes $\A_{IJ}^{(i)}$ and
$\B_{IJ}^{(i)}$ by the relations
\begin{eqnarray}
\A_{IJ}=-\sum_{i=1}^5
\xi_i\,\left(\gt\,{m_K^2-m_\pi^2\over12}\right)\,\A_{IJ}^{(i)},
\qquad
\B_{IJ}=-\sum_{i=1}^5 \xi_i\,\left(\gt\,{m_\pi^2\over4}\right)\,\B_{IJ}^{(i)}.
\label{defab}
\end{eqnarray}
     Here $\xi_i$ are parameters related to the Wilson coefficients
$c_i$ of (\ref{weak-lagr}) as
\begin{eqnarray}
&&\xi_1 = c_1 \left( 1 - {1 \over N_c} \right)\,, \quad
\xi_{2,3,4} = c_{2,3,4} \left( 1 + {1 \over N_c} \right)\,, \quad
\xi_5 = c_5 + {1 \over N_c} c_6 \,,
\end{eqnarray}
where the color factor ${1/N_c}$ originates from the Fierz-transformed
contribution to the nonleptonic weak effective chiral Lagrangian
\cite{bosoniz-our}.
    The normalization factors for the amplitudes $\A_{IJ}^{(i)}$ and
$\B_{IJ}^{(i)}$ in (\ref{defab}) were choosen in such way that in the
"soft pion" limit, corresponding to the $p^2$--order of the chiral
Lagrangian approach, one obtains
$$
\A_{11}^{(1)} = \B_{11}^{(1)} = -\A_{11}^{(2,3)} = -\B_{11}^{(2,3)} = -1.
$$
Then the other nonvanishing amplitudes in the soft pion limit are
$$
\A_{13}^{(4)} = 1\,,
\quad
\B_{13}^{(4)} = -\frac{1}{4} \, \frac{5 m_K^2 - 14 m_\pi^2}{m_K^2-m_\pi^2}\,,
\quad
\B_{23}^{(4)} =  \frac{9}{4} \frac{3 m_K^2 - 2 m_\pi^2}{m_K^2-m_\pi^2} \,.
$$

    The charge asymmetry of the slope parameters $\Delta g(K^\pm \to 3\pi)$
can be expressed by the formula
\begin{equation}
\Delta g \left( K^\pm \to
                          \left\{ {\Big.}^{\mbox{$\pi^\pm \pi^\pm \pi^\mp$}}
                                         _{\mbox{$\pi^0   \pi^0   \pi^\pm$}}
                          \right\}
         \right)
 = { \im F_1\, \sin(\alpha_1-\beta_1) \pm
     \im F_2\, \sin(\alpha_1-\beta_2) \over
     \re F_1\, \cos(\alpha_1-\beta_1) \pm
     \re F_2\, \cos(\alpha_1-\beta_2) } \,,
\label{delg}
\end{equation}
where $F_1=(a_{11}^* + a_{13}^*)(b_{11}+b_{13})$ and
$F_2=-(a_{11}^* + a_{13}^*)b_{23}$.
\footnote {In deriving charge asymmetries, one has to keep in mind,
that charge conjugation does reverse the phases of $\xi_i$ but not those
of $A_I^{(i)}$, $\A_{IJ}^{(i)}$, $\B_{IJ}^{(i)}$. }
   It is convenient to present the terms in the numerator of the
right-hand side of eq.(\ref{delg}) for $\delg$ in a more visual form
\begin{eqnarray}
\im F_1 &=& \Delta^{(1/2,\,1/2)} + \Delta^{(1/2,\,3/2)},
\nonumber \\
\Delta^{(1/2,\,1/2)} &=& \re a_{11}\,\im b_{11} - \im a_{11}\,\re b_{11}
\nonumber \\
    &=& \im \xi_5 \Big[ \re \B_{11}^{(5)} \big( \xi_{123} \re \A_{11}^{(1)}
                                               +\xi_5 \re \A_{11}^{(5)} \big)
                       -\re \A_{11}^{(5)} \big( \xi_{123} \re \B_{11}^{(1)}
                                               +\xi_5 \re \B_{11}^{(5)} \big)
                  \Big]\,,
\nonumber \\
\Delta^{(1/2,\,3/2)} &=& \re a_{13}\,\im b_{11} - \im a_{11}\,\re b_{13}
\nonumber \\
        &=& \xi_4 \, \im \xi_5 \big( \re \A_{13}^{(4)} \re \B_{11}^{(5)}
                                -\re \A_{11}^{(5)} \re \B_{13}^{(4)}\big)\,;
\nonumber \\
\im F_2 &=& \im a_{11} \, \re b_{23} \equiv \Delta^{'(1/2, \, 3/2)}
         = \xi_4 \, \im \xi_5 \re \A_{11}^{(5)} \re \B_{23}^{(4)} \,.
\label{delta}
\end{eqnarray}
   Here $\Delta^{(1/2, \, 1/2)}$ describes the contribution of the
interference of isotopic amplitudes $a_{11}$ and $b_{11}$ for transitions with
$|\Delta I|=1/2$, and $\Delta^{(1/2, \, 3/2)}$, $\Delta^{'(1/2, \, 3/2)}$ are
the contributions from interferences of amplitudes $a_{IJ}$ and $b_{IJ}$ with
$|\Delta I|=1/2$ and $3/2$.
   In writing eq.(\ref{delta}) we assume that direct CP-violation arises
only due to the imaginary parts of the isotopic amplitudes with
$|\Delta I|=1/2$ generated by the imaginary part of the Wilson coefficient
$c_5$ of the penguin operator $\O_5$.
   The fact that the relation
$\A_{11}^{(1)} = \B_{11}^{(1)} = -\A_{11}^{(2,3)} = -\B_{11}^{(2,3)}$
is fulfiled always if there is no isotopic symmetry breaking was also used.
   In this case the contribution of the nonpenguin operators with
$|\Delta I|=1/2$ to nonleptonic kaon decays can be joint to a term,
proportional to the combination $\xi_{123} = \big(-\xi_1+\xi_2+\xi_3 \big)$.

   In order to separate the contributions belonging to the
dominating combination $(-\xi_1+\xi_2+\xi_3)$ and to $\xi_4$,
$\xi_5$ respectively, we used in \cite{CPour-last} the experimental data on
$K\to2\pi$, $K\to3\pi$ decays and obtained the following values:
\begin{equation}
\xi_{123} =  6.96  \pm 0.48,  \qquad
\xi_4 =  0.516 \pm 0.025, \qquad \xi_5 = -0.183 \pm 0.022.
\label{xi}
\end{equation}
   As the analysis of the coefficients $c_i$ in leading--log
approximation of $QCD$ has shown, the main contribution to direct
$CP$--violation comes from the (gluonic) penguin diagram.
   The imaginary part of the coefficient $c_5$, responsible for the
direct $CP$--violation, can be fixed as \cite{CPour-last}
$$
   \im c_5 = 0.053^{+0.015}_{-0.011}|{\varepsilon}'/\varepsilon|\,.
$$

     The results of our calculations of $K \to 3\pi$ decay isotopic amplitudes
under successive inclusion of $p^4$--corrections are presented in table 1a
\footnote{The numerical values for the same isotopic amplitudes after
additional inclusion of $(\pi^0-\eta - \eta')$--mixing and one--loop
corrections were given in \cite{CPour-last}. }.
     The role of the $p^4$-contributions for the enhancement of
CP-violation due to interference of $|\Delta I| = 1/2$ amplitudes (the
quantity $\Delta^{(1/2,1/2)}$ in the expression for $\im F_1$ (\ref{delta}))
can be demonstrated by means of a simplified order of magnitude estimate of
the effect.
     In particular, taking only the $p^4$--corrections, arising
from the $L_{2,3}$--terms of Lagrangian (\ref{LeffGL}) and the currents
(\ref{JeffGL}, \ref{JSPnred}) (the Skyrme and non-Skyrme interactions), we
find for $a_{11}$ and $b_{11}$:
\begin{eqnarray*}
   a_{11} &=& \xi_{123} \Big(1+ \frac{m_K^2-3m_\pi^2}{12F_0^2\pi^2} \Big)
           +\xi_5\cdot 4R  \Big(1- \frac{m_K^2-3m_\pi^2}{12F_0^2\pi^2} \Big)\,,
\\
   b_{11} &=& \xi_{123} \Big(1- \frac{m_K^2+3m_\pi^2}{12F_0^2\pi^2} \Big)
           +\xi_5\cdot 4R  \Big(1+ \frac{m_K^2+3m_\pi^2}{12F_0^2\pi^2} \Big)\,.
\end{eqnarray*}
     Therefrom it is clear, that in the soft pion limit, with disappearence
of the $p^4$-contributions $\sim 1/(12F_0^2\pi^2)$, the contribution
from interference of $|\Delta I| = 1/2$ amplitudes equals zero.
     With $p^4$-corrections taken into account, we find for this
interference term the simple expression
\begin{eqnarray*}
\Delta^{(1/2,1/2)} = \im \xi_5 \, \re \xi_{123} \, 4R \frac{m_K^2}{3F_0^2\pi^2}
    \approx 149.1 \cdot \im c_5
\end{eqnarray*}
in good agreement with Table 1a.

    The Lagrangian (\ref{Lefflin}) and currents (\ref{JVAlin}), (\ref{JSPlin})
were used in \cite{shabalin1} for the calculation of $K \to 3\pi$ amplitudes
in the soft pion limit.
    In the last publication \cite{shabalin2} there was an attempt to
include in the linear $\sigma$--model also the $p^4$--corrections via
diagrams with intermediate scalar resonances.
    For the further discussion and comparison with the predictions of the
nonlinear Lagrangian approach it is convenient to present the results of the
calculations of $K \to 3\pi$ decay isotopic amplitudes of ref.\cite{shabalin2}
in a numerical form similar to Table 1a.
    The corresponding isotopic amplitudes in the soft pion limit and the
results of successive inclusion of $p^4$--corrections generated in the linear
$\sigma$--model by scalar resonance exchange are given in Table 1b.
    To present numerically the contributions of the penguin operator $\O_5$
we use the fact that the parameter
$\beta = 2m^4_{\pi}/\big[ (m_u + m_d)^2(m^2_{\sigma_{\pi}} - m^2_\pi)\big]
= 8.15$,
introduced in ref.\cite{shabalin2}, can be assumed to be equal in our notations
to the parameter $-R \approx 5.6$ defining the contribution of the
bosonized scalar density (\ref{JSPnred}).
\footnote{The (irrelevant) numerical discrepancy between $R$ and
$\beta$ disappears, if $(m_u+m_d)/2 = 6 MeV$ resp. $5MeV$ are taken
for $\beta$ resp. $R$ (in the text $5 MeV$ is used).}

    The quantities $\Delta^{(1/2, \, 1/2)}$, $\Delta^{(1/2, \, 3/2)}$ and
$\Delta^{'(1/2, \, 3/2)}$, corresponding to the definitions of
ref.\cite{shabalin2}, can be estimated after the replacement $\xi_i \to c_i$
in eqs.(\ref{delta}).
    Using the phenomenological relations for the Wilson coefficients of
ref.\cite{shabalin2}
\begin{eqnarray}
(c_1 - c_2 - c_3 - c_4) = -3.2 \,, \quad c_4 = 0.328 \,, \quad
(c_1 - c_2 - c_3 + 4\beta c_5) = -10.13
\label{cwilson}
\end{eqnarray}
one can fix the parameters $(c_1 - c_2 - c_3)$, $c_4$ and $4\beta c_5$
and obtaines for $\Delta^{(1/2, \, 1/2)}$, $\Delta^{(1/2, \, 3/2)}$ and
$\Delta^{'(1/2, \, 3/2)}$ the values which are also given in Table 1a.

    The comparison of Tables 1a and 1b shows that the
$p^4$--contributions owing to the $L_i$--terms of the nonlinear Lagrangian
(\ref{LeffGL}) and currents (\ref{JeffGL}), (\ref{JSPnred}) both
quantitatively and qualitatively differ from $p^4$--corrections
generated within linear $\sigma$--model by scalar resonance exchange.
    In the nonlinear Lagrangian approach the $p^4$--corrections increase
the amplitudes ${\cal A}^{(1)}_{11}$, ${\cal A}^{(4)}_{13}$,
${\cal B}^{(5)}_{11}$, ${\cal B}^{(4)}_{23}$ and decrease at the same
time the amplitudes ${\cal A}^{(5)}_{11}$, ${\cal B}^{(1)}_{11}$,
${\cal B}^{(4)}_{13}$ in their absolute values.
    On the other hand, the scalar resonance exchange increases the
absolute values of the amplitudes ${\cal A}^{(1)}_{11}$,
${\cal A}^{(4)}_{13}$, ${\cal B}^{(1, 5)}_{11}$, ${\cal B}^{(4)}_{13}$
and decreases the absolute values of the amplitudes
${\cal A}^{(5)}_{11}$, ${\cal B}^{(4)}_{23}$.
    Besides that, after taking into account the scalar resonance
exchange, the interferences $\Delta^{(1/2, \, 1/2)}$ and
$\Delta^{'(1/2, \, 3/2)}$ prove to be respectively by factors 4 and 2
less than in the nonlinear model.
    In consequence, not only the effect of the enhancement of direct
$CP$--violation by the interference $\Delta^{(1/2, \, 1/2)}$ is found to be
suppressed in the estimates of \cite{shabalin2} but also the
contribution of the interference $\Delta^{'(1/2, \, 3/2)}$,
nonvanishing in soft pion limit, is suppressed by a factor 2.

    The relative contribution of penguin and non--penguin operators in
the linear $\sigma$--model \cite{shabalin2} is determined by the ratio
$4\beta c_5/(c_1-c_2-c_3)=2.5$, showing, that about 80\% of the
observed amplitudes of $|\Delta I| = 1/2$ transitions in nonleptonic
kaon decays is attributed to the contribution of the penguin operator $\O_5$.
    On the other hand the parameters $\xi_i$ (\ref{xi}), fixed from
the analysis of $K \to 2\pi$ and $K \to 3\pi$ experimental data, lead
in the nonlinear chiral Lagrangian approach to the ratio
$4R\xi_5/\xi_{123}=0.58$.
    This ratio agrees with the results of our previous
phenomenological analysis \cite{CPour-prev}, where it was shown that
the contribution of the penguin operator is less $40\%$ of the
experimentally measured amplitudes of $|\Delta I| = 1/2$ transitions.
    These estimates confirm the results of a consistent analysis of
the $|\Delta I| = 1/2$ rule by Buras et al. \cite{buras1,buras2} which was
done in the nonlinear chiral Lagrangian approach with Wilson
coefficients calculated in leading--log approximation of QCD, where
the contribution of penguin operators to $K \to 2\pi$ decays was
estimated to be smaller then nonpenguin contribution within a wide
range of the renormalization scale $\mu$ (see also the analysis of
Wilson coefficients beyond leading logarithms in ref.\cite{CP-buras}).
    If one uses the parameters $\xi_i$ (\ref{xi}), instead of $c_i$
(\ref{cwilson}) to estimate the interference $\Delta^{(1/2,\,1/2)}$
in the linear $\sigma$--model, taking into account scalar resonance
exchange, the value $\Delta^{(1/2,\,1/2)}=108.5 \im c_5$ will be
obtained.
    This fact demonstrates that the estimates of the interference
$\Delta^{(1/2,\,1/2)}$ are very sensitive not only to the difference
of the dynamical behavior of penguin and non--penguin amplitudes at
$O(p^4)$ level but also to their relative contributions to $|\Delta I|
= 1/2$ transitions.
    In the case, when $|\Delta I| = 1/2$ transitions are dominated by the
contributions of the penguin operator, the interference
$\Delta^{(1/2,\,1/2)}$ becomes largely suppressed.

    In this way it appears, that in the framework of \cite{shabalin2}
the interference term $\Delta^{(1/2,\,1/2)}$, which is mainly
responsible for the enhancement of the charge asymmetry, is lower by a
factor $\sim 4$ (as compared to \cite{CPour-last}) already in Born
approximation.
    The reason for this discrepancy has been discussed in some detail
above.
    There is an additional enhancement of $\Delta^{(1/2,\,1/2)}$ by a
factor 3 from contributions of meson loops to the real parts
of isotopic amplitudes (see \cite{CPour-last}).
    (In \cite{shabalin2} only the absorptive parts of meson loops have
been calculated).
    As a result, our value for the charge asymmetry $\Delta g$ in
$K^{\pm} \to 3\pi$ decays \cite{CPour-last} should be about
12 times larger then that estimated by \cite{shabalin2},
whereby the discrepancy can be explained by the fact, that in the
latter case by using a linear $\sigma$--model other corrections of
order $p^4$ are considered, and by a more complete treatment of loop
corrections in the first case.
\footnote{The difference of the estimates of ref.\cite{CPour-last}
with respect to the one given earlier \cite{CPour-prev} (less than a
factor 2) is due to the effect of $(\pi^0, \eta, \eta')$--mixing, formely
not taken into account completely.
    On the other hand this suppression is compensated by a larger mass of
the $t$--quark $m_t \geq 100 GeV$ by the additional (relative)
enhancement effect arising from contributions of the electroweak
penguin operator $\O_8$.}

    We leave aside a detailed discussion of the (strong) phase
differences between isotopic amplitudes, which appear after the
calculation of meson loops.
    As they are determined by the ratios of imaginary and real parts
of amplitudes, a modification of the latter may be important and
should be taken into account.
    In our case the phases have been extracted from direct
calculations of one loop diagrams, using superpropagator
regularization.
    Results of analogous calculations for $K \to 2\pi$ decays
\cite{K2pi-our} are in agreement with Kambor et al. \cite{kambor},
where for the regularization of $UV$--divergences the usual method of
introducing counter terms into the Lagrangian was used.
     We should mention, that in the latter paper also sizeable imaginary
parts for the amplitudes $\beta$ (see table 2, loc. cit.) are found by
the loop calculation, but the resulting phases are then suppressed by
the choice of counterterms, making their perturbative approach --
as the authors themselves remark -- somewhat problematic.
     Of course it would be interesting to fix them directly from
experimental data, but this was not possible until now neither in our fit
nor in other work.

    Concerning the linear $\sigma$--model in general, it was
demonstrated by Gasser and Leutwyler \cite{Gasser-Leutwyler}, that it
has the correct chiral structure, but a wrong phenomenology at the
next-to-leading order in the chiral expansion for any value of the
scalar resonance mass (see also the criticism of the linear
$\sigma$--model by Meissner \cite{meissner} and lectures by Ecker
\cite{ecker-lecture} and Pich \cite{pich-lecture}).
    The $p^4$--corrections generated by
scalar resonance exchange in the linear $\sigma$--model are not
equivalent to the $p^4$--corrections related to the $L_i$ terms of the
nonlinear Lagrangian (\ref{LeffGL}) and the currents (\ref{JeffGL}),
(\ref{JSPnred}).
    It was shown by Ecker et al. \cite{ecker-strongres} and by
Donoghue et al. \cite{donoghue-strongres} that the structure constants
$L_i$ of the effective chiral Lagrangian for strong interactions of
order $p^4$ are largely saturated by vector resonance exchange at order $p^2$.
    The most general analysis has been carried out in
ref.\cite{ecker-strongres}, where all possible chiral couplings to the
pseudoscalar mesons linear in the resonance fields were constructed to
lowest order in the chiral expansion.
    In particular, the coupling constants $L_{2,3}$ are completely dominated
by vector resonance exchange while scalar resonances contribute only
to $L_3$, and this contribution does not exceed 20\%.
    The derivation of the Skyrme--type $p^4$--interaction from the integrated
out vector and axial--vector resonances has been given also by
Igarashi \cite{igarashi} from the hidden--local--symmetry Lagrangian of
Bando et al. \cite{bando}.
    The resonance contributions to the pseudoscalar weak Lagrangian and the
modification of its structure after integrating out the heavy meson exchanges
were discussed recently by Ecker et al. \cite{ecker-weakres} and by
Isidori and Pugliese \cite{isidori-weakres}.
    Thus, the attempt of ref.\cite{shabalin2} to reproduce the
$p^4$--contributions of the nonlinear Lagrangian (\ref{LeffGL}) and
the currents (\ref{JeffGL}), (\ref{JSPnred}) within the linear
$\sigma$--model by taking into account intermediate scalar
resonances seems to be not justified also for other phenomenological
reasons.

   The authors are grateful to A.J.Buras, G.Ecker, J.Gasser,
A.Pugliese and G.Isidory for useful discussions and helpful comments.
   One of the authors (A.A.Bel'kov) thanks the Institute of Elementary
Particle Physics, Humboldt-University, Berlin and DESY--Institute for
High Energy Physics, Zeuthen for their hospitality.
He acknowledges the support from DFG, Project Eb 139/1--1.
   A.V.Lanyov is grateful for the hospitality extended to him at the
DESY--Institute for High Energy Physics, Zeuthen.


\newpage


%
%

\begin{tabular}{lp{10cm}}
\underline{Table 1} & Comparison of the $K \rightarrow 3 \pi$
                      isotopic amplitudes, calculated in non--linear
                      and linear chiral Lagrangian approaches
                      (Born approximation)\\
\end{tabular}

\indent

\smallskip
a)  Nonlinear chiral Lagrangian approach \cite{CPour-last}\\
\medskip

\begin{tabular}{|c|ccc|ccc|} \hline
&\multicolumn{3}{c|}{Soft--pion limit}&\multicolumn{3}{c|}{Inclusion of p$^4$
--corrections }\\ \cline{2-7}
\raisebox{1.5ex}[-1.5ex]{}&$\O_1$ &$\O_4$ &$\O_5$ &$\O_1$ &$\O_4$ &$\O_5$\\
[0.5ex]\hline
$\re \A_{11}$ & -1.00 & & -22.42& -1.22 & & -19.22 \\
$\re \B_{11}$ & -1.00 & & -22.42 & -0.70 & & -30.45 \\
$\re \A_{13}$ &  & 1.00 &  & & 1.22 & \\
$\re \B_{13}$ & & -1.07 & & & -0.91 &  \\
$\re \B_{23}$ & & 6.93 & & & 7.39 & \\ \hline
$\Delta^{(1/2,1/2)}$ & & 0 & & & 165.4 Im $c_5$ & \\
$\Delta^{(1/2,3/2)}$ & & 22.5 Im $c_5$ & & & 26.4 Im $c_5$ & \\
$\Delta^{'(1/2, 3/2)}$ & & 74.9 Im $c_5$ & & & 73.2 Im $c_5$ & \\ \hline
\end{tabular}
\vspace*{1.0cm}

b) Linear $\sigma$--model \cite{shabalin2}\\
\medskip

\begin{tabular}{|c|ccc|ccc|} \hline
&\multicolumn{3}{c|}{Soft--pion limit}&\multicolumn{3}{c|}{Inclusion of p$^4$
--corrections }\\ \cline{2-7}
\raisebox{1.5ex}[-1.5ex]{}&$\O_1$ &$\O_4$ &$\O_5$ &$\O_1$ &$\O_4$ &$\O_5$\\
[0.5ex]\hline
$\re \A_{11}$ & -1.00 & & -22.42 & -1.80 & & -19.28 \\
$\re \B_{11}$ & -1.00 & & -22.42 & -1.63 & & -26.19 \\
$\re \A_{13}$ &  & 1.00 &  & & 1.80 & \\
$\re \B_{13}$ & & -1.25 & & & -1.78 & \\
$\re \B_{23}$ & & 6.75 & & & 6.22 & \\  \hline
$\Delta^{(1/2,1/2)}$ & & 0 & & & 45.6 Im $c_5$ & \\
$\Delta^{(1/2,3/2)}$ & & 15.2 Im $c_5$ & & & 26.7 Im $c_5$ & \\
$\Delta^{'(1/2, 3/2)}$ & & 49.6 Im $c_5$ & & & 39.3 Im $c_5$ & \\ \hline
\end{tabular}
\end{document}